\documentclass[aps,twocolumn,superscriptaddress,showpacs]{revtex4}

\usepackage{epsfig}
\usepackage{graphicx}
\usepackage{dcolumn}
\usepackage{amsmath}
\usepackage{xspace}
\usepackage{setspace}
\usepackage{ifthen}

\usepackage[font=small,labelsep=period,justification=raggedright]{caption}

\def \Et {{E}_{T}}

\newcommand{\pt}{{ p}_{ T}}

\newcommand{\met}{\mbox{${ \not\! E}_{ T}$}}
\newcommand{\Met}{\mbox{${ \not\! E}_{ T}$}} 
\newcommand{\metvec}{{\not\!\! \vec{E}_T}}

\def \Ht {{H}_{T}}

\newcommand{\Zgstar}{Z^0\kern -0.25em/\kern -0.15em\gamma^*}
\def\Zgamma{\Z\kern -0.1em/\kern -0.1em\gamma}

\def\lgmet{\ell\gamma\Met}

\def \lmetb{\ell\Met b}

\def\llg{\ell\ell\gamma}

\def\eemet{ee\Met}

\def\Wg{W\gamma}

\def\deg{^\circ}

\def\GeV{\rm GeV}

\def\GeVc2{GeV/{c^2}}

\newcommand{\invfb}{\rm fb^{-1}}

\newcommand{\ppbar}{p{\bar p}} 

\def\ccbar{c{\bar c}}
\def\bbbar{b{\bar b}}

\def\ttbar{t\bar{t}}

\def\ttg{t\bar{t}\gamma}

\newcommand{\goes}{\kern -0.18em\rightarrow\kern -0.18em}
\newcommand{\plus}{\kern -0.18em +\kern -0.18em}
\def\Z{Z^0}

\def\roots{{\sqrt s}}

\def\95cl{95 \%~C.L.}
\def\95CL{95 \%~C.L.}
\def\r#1 {$^{#1}$}

\newcommand{\pmasym}[2]{^{\mbox{\small +#1}}_{\hskip 1pt\mbox{\small --#2}}}
\newcounter{myquestion}

\newcommand{\TeV}{\ensuremath{\mathrm{Te\kern -0.1em V}}\xspace}
\newcommand{\bfTeV}{\ensuremath{\bf{Te\kern -0.1em V}}\xspace}
\newcommand{\rrr}{\rightarrow}

\newcommand{\Njets}{\ensuremath{\mathrm{N_{jets}}}}
\newcommand{\Etgamma}{{E_T^{\gamma}}}
\newcommand{\Etjet}{E_T^{jet}}

\newcommand{\Etlepton}{E_T^{\ell}}

\newcommand{\apprluminosityfb}{6.0}
\newcommand{\ratiottgttbar}{0.024}
\newcommand{\dratiottgttbar}{0.009}

\newcommand{\totxsecnoflttbargcutsisojetssuper}{0.18}
\newcommand{\totxsecdnoflttbargcutsisojetssuper}{0.07}
\newcommand{\totxsecdsysnoflttbargcutsisojetssuper}{0.04}
\newcommand{\totxsecdlumnoflttbargcutsisojetssuper}{0.01}
\newcommand{\totxsecdtotnoflttbargcutsisojetssuper}{0.08}

\newcommand{\smdtotnofettbargcutsisojetssuper}{2.2}

\newcommand{\smdtotnoflttbargcutsisojetssuper}{3.4}

\newcommand{\smdtotnofmuttbargcutsisojetssuper}{1.9}

\newcommand{\totefpdsysnofettbargcutsisojetssuper}{0.19}

\newcommand{\totefpdsysnofmuttbargcutsisojetssuper}{0.11}

\newcommand{\totjfakepdsysnofettbargcutsisojetssuper}{1.76}

\newcommand{\totjfakepdsysnoflttbargcutsisojetssuper}{2.53}

\newcommand{\totjfakepdsysnofmuttbargcutsisojetssuper}{1.56}

\newcommand{\totjqcddsysnofettbargcutsisojetssuper}{0.38}

\newcommand{\totjqcddsysnoflttbargcutsisojetssuper}{0.38}

\newcommand{\totjqcddsysnofmuttbargcutsisojetssuper}{0.02}

\newcommand{\totmistagdsysnofettbargcutsisojetssuper}{0.37}

\newcommand{\totmistagdsysnoflttbargcutsisojetssuper}{0.51}

\newcommand{\totmistagdsysnofmuttbargcutsisojetssuper}{0.32}

\newcommand{\totstopsdsysnofettbargcutsisojetssuper}{0.10}

\newcommand{\totstopsdsysnoflttbargcutsisojetssuper}{0.13}

\newcommand{\totstopsdsysnofmuttbargcutsisojetssuper}{0.10}

\newcommand{\totstoptdsysnofettbargcutsisojetssuper}{0.14}

\newcommand{\totstoptdsysnoflttbargcutsisojetssuper}{0.19}

\newcommand{\totstoptdsysnofmuttbargcutsisojetssuper}{0.14}

\newcommand{\tottaufakeedsysnofettbargcutsisojetssuper}{0.08}

\newcommand{\tottaufakeedsysnoflttbargcutsisojetssuper}{0.09}

\newcommand{\tottaufakeedsysnofmuttbargcutsisojetssuper}{0.05}

\newcommand{\totttadildsysnofettbargcutsisojetssuper}{0.27}

\newcommand{\totttadildsysnoflttbargcutsisojetssuper}{0.50}

\newcommand{\totttadildsysnofmuttbargcutsisojetssuper}{0.24}

\newcommand{\totttasemdsysnofettbargcutsisojetssuper}{1.10}

\newcommand{\totttasemdsysnoflttbargcutsisojetssuper}{2.04}

\newcommand{\totttasemdsysnofmuttbargcutsisojetssuper}{0.97}

\newcommand{\totwbbadsysnofettbargcutsisojetssuper}{0.07}

\newcommand{\totwbbadsysnoflttbargcutsisojetssuper}{0.08}

\newcommand{\totwbbadsysnofmuttbargcutsisojetssuper}{0.05}

\newcommand{\totwcadsysnofettbargcutsisojetssuper}{0.07}

\newcommand{\totwcadsysnoflttbargcutsisojetssuper}{0.09}

\newcommand{\totwcadsysnofmuttbargcutsisojetssuper}{0.07}

\newcommand{\totwccadsysnofettbargcutsisojetssuper}{0.05}

\newcommand{\totwccadsysnoflttbargcutsisojetssuper}{0.07}

\newcommand{\totwccadsysnofmuttbargcutsisojetssuper}{0.05}

\newcommand{\totwwdsysnofettbargcutsisojetssuper}{0.03}

\newcommand{\totwwdsysnoflttbargcutsisojetssuper}{0.03}

\newcommand{\totwwdsysnofmuttbargcutsisojetssuper}{0.03}

\newcommand{\totwzdsysnofettbargcutsisojetssuper}{0.05}

\newcommand{\totwzdsysnoflttbargcutsisojetssuper}{0.06}

\newcommand{\totwzdsysnofmuttbargcutsisojetssuper}{0.05}

\newcommand{\stopsnoflttbargcutsisojetssuper}{0.09}

\newcommand{\stoptnoflttbargcutsisojetssuper}{0.27}

\newcommand{\taufakeenoflttbargcutsisojetssuper}{0.29}

\newcommand{\ttasemnoflttbargcutsisojetssuper}{11.19}

\newcommand{\ttadilnoflttbargcutsisojetssuper}{2.74}

\newcommand{\wbbanoflttbargcutsisojetssuper}{0.21}

\newcommand{\wcanoflttbargcutsisojetssuper}{0}

\newcommand{\wccanoflttbargcutsisojetssuper}{0.05}

\newcommand{\wznoflttbargcutsisojetssuper}{0.09}

\newcommand{\wwnoflttbargcutsisojetssuper}{0.11}

\newcommand{\stopsnofettbargcutsisojetssuper}{0.09}

\newcommand{\stoptnofettbargcutsisojetssuper}{0.14}

\newcommand{\taufakeenofettbargcutsisojetssuper}{0.20}

\newcommand{\ttasemnofettbargcutsisojetssuper}{5.98}

\newcommand{\ttadilnofettbargcutsisojetssuper}{1.47}

\newcommand{\wbbanofettbargcutsisojetssuper}{0.15}

\newcommand{\wcanofettbargcutsisojetssuper}{0}

\newcommand{\wccanofettbargcutsisojetssuper}{0}

\newcommand{\wznofettbargcutsisojetssuper}{0.05}

\newcommand{\wwnofettbargcutsisojetssuper}{0.06}

\newcommand{\stopsnofmuttbargcutsisojetssuper}{0}

\newcommand{\stoptnofmuttbargcutsisojetssuper}{0.13}

\newcommand{\taufakeenofmuttbargcutsisojetssuper}{0.10}

\newcommand{\ttasemnofmuttbargcutsisojetssuper}{5.21}

\newcommand{\ttadilnofmuttbargcutsisojetssuper}{1.27}

\newcommand{\wbbanofmuttbargcutsisojetssuper}{0.06}

\newcommand{\wcanofmuttbargcutsisojetssuper}{0}

\newcommand{\wccanofmuttbargcutsisojetssuper}{0.05}

\newcommand{\wznofmuttbargcutsisojetssuper}{0.05}

\newcommand{\wwnofmuttbargcutsisojetssuper}{0.06}

\newcommand{\datanofmuttbargcutsisojetssuper}{13}

\newcommand{\datanofettbargcutsisojetssuper}{17}

\newcommand{\datanoflttbargcutsisojetssuper}{30}

\newcommand{\corrjfakepnoflttbargcutsisojetssuper}{7.54}

\newcommand{\corrmistagnoflttbargcutsisojetssuper}{2.50}
\newcommand{\corrjqcdnoflttbargcutsisojetssuper}{0.40}
\newcommand{\corrjfakepnofmuttbargcutsisojetssuper}{1.79}
\newcommand{\correfpnofmuttbargcutsisojetssuper}{0.49}
\newcommand{\corrmistagnofmuttbargcutsisojetssuper}{1.02}
\newcommand{\corrjqcdnofmuttbargcutsisojetssuper}{0.02}
\newcommand{\corrjfakepnofettbargcutsisojetssuper}{5.75}
\newcommand{\correfpnofettbargcutsisojetssuper}{0.94}
\newcommand{\corrmistagnofettbargcutsisojetssuper}{1.47}
\newcommand{\corrjqcdnofettbargcutsisojetssuper}{0.38}

\newcommand{\smnodcnofettbargcutsisojetssuper}{16.7}
\newcommand{\smnodcnofmuttbargcutsisojetssuper}{10.3}
\newcommand{\smnodcnoflttbargcutsisojetssuper}{26.9}

\newcommand{\xsecnoflttbarcutsisojets}{7.62}
\newcommand{\xsecdnoflttbarcutsisojets}{0.20}
\newcommand{\xsecdsysnoflttbarcutsisojets}{0.68}
\newcommand{\xsecdlumnoflttbarcutsisojets}{0.46}

\newcommand{\smdtotnofettbarcutsisojets}{196}
\newcommand{\smdtotnoflttbarcutsisojets}{340}
\newcommand{\smdtotnofmuttbarcutsisojets}{146}

\newcommand{\smnoflttbarcutsisojets}{4420}

\newcommand{\totjqcddsysnofettbarcutsisojets}{38}
\newcommand{\totjqcddsysnoflttbarcutsisojets}{40}
\newcommand{\totjqcddsysnofmuttbarcutsisojets}{3}
\newcommand{\totmistagdsysnofettbarcutsisojets}{29}
\newcommand{\totmistagdsysnoflttbarcutsisojets}{46}
\newcommand{\totmistagdsysnofmuttbarcutsisojets}{17}
\newcommand{\totstopsdsysnofettbarcutsisojets}{10}
\newcommand{\totstopsdsysnoflttbarcutsisojets}{18}
\newcommand{\totstopsdsysnofmuttbarcutsisojets}{8}
\newcommand{\totstoptdsysnofettbarcutsisojets}{9}
\newcommand{\totstoptdsysnoflttbarcutsisojets}{16}
\newcommand{\totstoptdsysnofmuttbarcutsisojets}{7}
\newcommand{\totttdsysnofettbarcutsisojets}{180}
\newcommand{\totttdsysnoflttbarcutsisojets}{330}
\newcommand{\totttdsysnofmuttbarcutsisojets}{140}

\newcommand{\totwwdsysnofettbarcutsisojets}{4}
\newcommand{\totwwdsysnoflttbarcutsisojets}{7}
\newcommand{\totwwdsysnofmuttbarcutsisojets}{3}
\newcommand{\totwznewdsysnofettbarcutsisojets}{1.1}
\newcommand{\totwznewdsysnoflttbarcutsisojets}{2.0}
\newcommand{\totwznewdsysnofmuttbarcutsisojets}{0.9}

\newcommand{\totzzdsysnofettbarcutsisojets}{0.2}
\newcommand{\totzzdsysnoflttbarcutsisojets}{0.3}
\newcommand{\totzzdsysnofmuttbarcutsisojets}{0.1}

\newcommand{\wcincdtotnofettbarcutsisojets}{13}
\newcommand{\wcincdtotnoflttbarcutsisojets}{23}
\newcommand{\wcincdtotnofmuttbarcutsisojets}{9}
\newcommand{\wcincnofettbarcutsisojets}{85}
\newcommand{\wcincnoflttbarcutsisojets}{147}
\newcommand{\wcincnofmuttbarcutsisojets}{61}

\newcommand{\wccincdtotnofettbarcutsisojets}{23}
\newcommand{\wccincdtotnoflttbarcutsisojets}{40}
\newcommand{\wccincdtotnofmuttbarcutsisojets}{17}
\newcommand{\wccincnofettbarcutsisojets}{127}
\newcommand{\wccincnoflttbarcutsisojets}{221}
\newcommand{\wccincnofmuttbarcutsisojets}{94}

\newcommand{\wbbincdtotnofettbarcutsisojets}{34}
\newcommand{\wbbincdtotnoflttbarcutsisojets}{58}
\newcommand{\wbbincdtotnofmuttbarcutsisojets}{24}
\newcommand{\wbbincnofettbarcutsisojets}{203}
\newcommand{\wbbincnoflttbarcutsisojets}{348}
\newcommand{\wbbincnofmuttbarcutsisojets}{146}

\newcommand{\zbbincdtotnofettbarcutsisojets}{3}
\newcommand{\zbbincdtotnoflttbarcutsisojets}{5}
\newcommand{\zbbincdtotnofmuttbarcutsisojets}{2}
\newcommand{\zbbincnofettbarcutsisojets}{31}
\newcommand{\zbbincnoflttbarcutsisojets}{53}
\newcommand{\zbbincnofmuttbarcutsisojets}{22}

\newcommand{\zttincdtotnofettbarcutsisojets}{8}
\newcommand{\zttincdtotnoflttbarcutsisojets}{11}
\newcommand{\zttincdtotnofmuttbarcutsisojets}{8}
\newcommand{\zttincnofettbarcutsisojets}{6}
\newcommand{\zttincnoflttbarcutsisojets}{14}
\newcommand{\zttincnofmuttbarcutsisojets}{9}

\newcommand{\stopsnoflttbarcutsisojets}{135}

\newcommand{\stoptnoflttbarcutsisojets}{116}

\newcommand{\ttnoflttbarcutsisojets}{2500}

\newcommand{\wwnoflttbarcutsisojets}{51}

\newcommand{\wznewnoflttbarcutsisojets}{15.1}

\newcommand{\zznoflttbarcutsisojets}{2.3}

\newcommand{\stopsnofettbarcutsisojets}{76}

\newcommand{\stoptnofettbarcutsisojets}{66}

\newcommand{\ttnofettbarcutsisojets}{1420}

\newcommand{\wwnofettbarcutsisojets}{29}

\newcommand{\wznewnofettbarcutsisojets}{8.6}

\newcommand{\zznofettbarcutsisojets}{1.3}

\newcommand{\stopsnofmuttbarcutsisojets}{59}

\newcommand{\stoptnofmuttbarcutsisojets}{50}

\newcommand{\ttnofmuttbarcutsisojets}{1080}

\newcommand{\wwnofmuttbarcutsisojets}{22}

\newcommand{\wznewnofmuttbarcutsisojets}{6.5}

\newcommand{\zznofmuttbarcutsisojets}{1.0}

\newcommand{\jqcdnofettbarcutsisojets}{222}

\newcommand{\jqcdnofmuttbarcutsisojets}{20}

\newcommand{\jqcdnoflttbarcutsisojets}{240}

\newcommand{\mistagnofettbarcutsisojets}{358}

\newcommand{\mistagnofmuttbarcutsisojets}{214}

\newcommand{\mistagnoflttbarcutsisojets}{572}

\newcommand{\datanofmuttbarcutsisojets}{1709}
\newcommand{\datanofettbarcutsisojets}{2720}
\newcommand{\datanoflttbarcutsisojets}{4429}

\newcommand{\smnodcnofettbarcutsisojets}{2630}
\newcommand{\smnodcnofmuttbarcutsisojets}{1790}
\newcommand{\smnodcnoflttbarcutsisojets}{4420}

\newcommand{\datanoflwgiso}{8276}

\begin{document}

\title{Evidence for $\ttg$ Production and Measurement of $\sigma_{\ttg}/\sigma_{\ttbar}$ }
\affiliation{Institute of Physics, Academia Sinica, Taipei, Taiwan 11529, Republic of China} 
\affiliation{Argonne National Laboratory, Argonne, Illinois 60439, USA} 
\affiliation{University of Athens, 157 71 Athens, Greece} 
\affiliation{Institut de Fisica d'Altes Energies, ICREA, Universitat Autonoma de Barcelona, E-08193, Bellaterra (Barcelona), Spain} 
\affiliation{Baylor University, Waco, Texas 76798, USA} 
\affiliation{Istituto Nazionale di Fisica Nucleare Bologna, $^z$University of Bologna, I-40127 Bologna, Italy} 
\affiliation{University of California, Davis, Davis, California 95616, USA} 
\affiliation{University of California, Los Angeles, Los Angeles, California 90024, USA} 
\affiliation{Instituto de Fisica de Cantabria, CSIC-University of Cantabria, 39005 Santander, Spain} 
\affiliation{Carnegie Mellon University, Pittsburgh, Pennsylvania 15213, USA} 
\affiliation{Enrico Fermi Institute, University of Chicago, Chicago, Illinois 60637, USA}
\affiliation{Comenius University, 842 48 Bratislava, Slovakia; Institute of Experimental Physics, 040 01 Kosice, Slovakia} 
\affiliation{Joint Institute for Nuclear Research, RU-141980 Dubna, Russia} 
\affiliation{Duke University, Durham, North Carolina 27708, USA} 
\affiliation{Fermi National Accelerator Laboratory, Batavia, Illinois 60510, USA} 
\affiliation{University of Florida, Gainesville, Florida 32611, USA} 
\affiliation{Laboratori Nazionali di Frascati, Istituto Nazionale di Fisica Nucleare, I-00044 Frascati, Italy} 
\affiliation{University of Geneva, CH-1211 Geneva 4, Switzerland} 
\affiliation{Glasgow University, Glasgow G12 8QQ, United Kingdom} 
\affiliation{Harvard University, Cambridge, Massachusetts 02138, USA} 
\affiliation{Division of High Energy Physics, Department of Physics, University of Helsinki and Helsinki Institute of Physics, FIN-00014, Helsinki, Finland} 
\affiliation{University of Illinois, Urbana, Illinois 61801, USA} 
\affiliation{The Johns Hopkins University, Baltimore, Maryland 21218, USA} 
\affiliation{Institut f\"{u}r Experimentelle Kernphysik, Karlsruhe Institute of Technology, D-76131 Karlsruhe, Germany} 
\affiliation{Center for High Energy Physics: Kyungpook National University, Daegu 702-701, Korea; Seoul National University, Seoul 151-742, Korea; Sungkyunkwan University, Suwon 440-746, Korea; Korea Institute of Science and Technology Information, Daejeon 305-806, Korea; Chonnam National University, Gwangju 500-757, Korea; Chonbuk National University, Jeonju 561-756, Korea} 
\affiliation{Ernest Orlando Lawrence Berkeley National Laboratory, Berkeley, California 94720, USA} 
\affiliation{University of Liverpool, Liverpool L69 7ZE, United Kingdom} 
\affiliation{University College London, London WC1E 6BT, United Kingdom} 
\affiliation{Centro de Investigaciones Energeticas Medioambientales y Tecnologicas, E-28040 Madrid, Spain} 
\affiliation{Massachusetts Institute of Technology, Cambridge, Massachusetts 02139, USA} 
\affiliation{Institute of Particle Physics: McGill University, Montr\'{e}al, Qu\'{e}bec, Canada H3A~2T8; Simon Fraser University, Burnaby, British Columbia, Canada V5A~1S6; University of Toronto, Toronto, Ontario, Canada M5S~1A7; and TRIUMF, Vancouver, British Columbia, Canada V6T~2A3} 
\affiliation{University of Michigan, Ann Arbor, Michigan 48109, USA} 
\affiliation{Michigan State University, East Lansing, Michigan 48824, USA}
\affiliation{Institution for Theoretical and Experimental Physics, ITEP, Moscow 117259, Russia}
\affiliation{University of New Mexico, Albuquerque, New Mexico 87131, USA} 
\affiliation{Northwestern University, Evanston, Illinois 60208, USA} 
\affiliation{The Ohio State University, Columbus, Ohio 43210, USA} 
\affiliation{Okayama University, Okayama 700-8530, Japan} 
\affiliation{Osaka City University, Osaka 588, Japan} 
\affiliation{University of Oxford, Oxford OX1 3RH, United Kingdom} 
\affiliation{Istituto Nazionale di Fisica Nucleare, Sezione di Padova-Trento, $^{aa}$University of Padova, I-35131 Padova, Italy} 
\affiliation{LPNHE, Universite Pierre et Marie Curie/IN2P3-CNRS, UMR7585, Paris, F-75252 France} 
\affiliation{University of Pennsylvania, Philadelphia, Pennsylvania 19104, USA}
\affiliation{Istituto Nazionale di Fisica Nucleare Pisa, $^{bb}$University of Pisa, $^{cc}$University of Siena and $^{dd}$Scuola Normale Superiore, I-56127 Pisa, Italy} 
\affiliation{University of Pittsburgh, Pittsburgh, Pennsylvania 15260, USA} 
\affiliation{Purdue University, West Lafayette, Indiana 47907, USA} 
\affiliation{University of Rochester, Rochester, New York 14627, USA} 
\affiliation{The Rockefeller University, New York, New York 10065, USA} 
\affiliation{Istituto Nazionale di Fisica Nucleare, Sezione di Roma 1, $^{ee}$Sapienza Universit\`{a} di Roma, I-00185 Roma, Italy} 

\affiliation{Rutgers University, Piscataway, New Jersey 08855, USA} 
\affiliation{Texas A\&M University, College Station, Texas 77843, USA} 
\affiliation{Istituto Nazionale di Fisica Nucleare Trieste/Udine, I-34100 Trieste, $^{ff}$University of Trieste/Udine, I-33100 Udine, Italy} 
\affiliation{University of Tsukuba, Tsukuba, Ibaraki 305, Japan} 
\affiliation{Tufts University, Medford, Massachusetts 02155, USA} 
\affiliation{University of Virginia, Charlottesville, VA  22906, USA}
\affiliation{Waseda University, Tokyo 169, Japan} 
\affiliation{Wayne State University, Detroit, Michigan 48201, USA} 
\affiliation{University of Wisconsin, Madison, Wisconsin 53706, USA} 
\affiliation{Yale University, New Haven, Connecticut 06520, USA} 
\author{T.~Aaltonen}
\affiliation{Division of High Energy Physics, Department of Physics, University of Helsinki and Helsinki Institute of Physics, FIN-00014, Helsinki, Finland}
\author{B.~\'{A}lvarez~Gonz\'{a}lez$^v$}
\affiliation{Instituto de Fisica de Cantabria, CSIC-University of Cantabria, 39005 Santander, Spain}
\author{S.~Amerio}
\affiliation{Istituto Nazionale di Fisica Nucleare, Sezione di Padova-Trento, $^{aa}$University of Padova, I-35131 Padova, Italy} 

\author{D.~Amidei}
\affiliation{University of Michigan, Ann Arbor, Michigan 48109, USA}
\author{A.~Anastassov}
\affiliation{Northwestern University, Evanston, Illinois 60208, USA}
\author{A.~Annovi}
\affiliation{Laboratori Nazionali di Frascati, Istituto Nazionale di Fisica Nucleare, I-00044 Frascati, Italy}
\author{J.~Antos}
\affiliation{Comenius University, 842 48 Bratislava, Slovakia; Institute of Experimental Physics, 040 01 Kosice, Slovakia}
\author{G.~Apollinari}
\affiliation{Fermi National Accelerator Laboratory, Batavia, Illinois 60510, USA}
\author{J.A.~Appel}
\affiliation{Fermi National Accelerator Laboratory, Batavia, Illinois 60510, USA}
\author{A.~Apresyan}
\affiliation{Purdue University, West Lafayette, Indiana 47907, USA}
% ...
\collaboration{CDF Collaboration\footnote{With visitors from $^a$University of Massachusetts Amherst, Amherst, Massachusetts 01003,
$^b$Istituto Nazionale di Fisica Nucleare, Sezione di Cagliari, 09042 Monserrato (Cagliari), Italy,
$^c$University of California Irvine, Irvine, CA  92697, 
$^d$University of California Santa Barbara, Santa Barbara, CA 93106
$^e$University of California Santa Cruz, Santa Cruz, CA  95064,
$^f$CERN,CH-1211 Geneva, Switzerland,
$^g$Cornell University, Ithaca, NY  14853, 
$^h$University of Cyprus, Nicosia CY-1678, Cyprus, 
$^i$University College Dublin, Dublin 4, Ireland,
$^j$University of Fukui, Fukui City, Fukui Prefecture, Japan 910-0017,
$^k$Universidad Iberoamericana, Mexico D.F., Mexico,
$^l$Iowa State University, Ames, IA  50011,
$^m$University of Iowa, Iowa City, IA  52242,
$^n$Kinki University, Higashi-Osaka City, Japan 577-8502,
$^o$Kansas State University, Manhattan, KS 66506,
$^p$University of Manchester, Manchester M13 9PL, England,
$^q$Queen Mary, University of London, London, E1 4NS, England,
$^r$Muons, Inc., Batavia, IL 60510,
$^s$Nagasaki Institute of Applied Science, Nagasaki, Japan, 
$^t$National Research Nuclear University, Moscow, Russia,
$^u$University of Notre Dame, Notre Dame, IN 46556,
$^v$Universidad de Oviedo, E-33007 Oviedo, Spain, 
$^w$Texas Tech University, Lubbock, TX  79609, 
$^x$Universidad Tecnica Federico Santa Maria, 110v Valparaiso, Chile,
$^y$Yarmouk University, Irbid 211-63, Jordan,
$^{gg}$On leave from J.~Stefan Institute, Ljubljana, Slovenia, 
}}
\noaffiliation

\preprint{FERMILAB-PUB-07/yyy-E}

\date{\today}

\begin{abstract}
  Using data corresponding to $\apprluminosityfb~\invfb$ of $\ppbar$
  collisions at $\sqrt{s}=1.96$ TeV collected by the CDF II detector,
  we present a cross section measurement of top-quark pair production
  with an additional radiated photon, $\ttg$. The events are selected
  by looking for a lepton ($\ell$), a photon ($\gamma$), significant
  transverse momentum imbalance ($\met$), large total transverse
  energy, and three or more jets, with at least one identified as
  containing a $b$ quark ($b$).  The $\ttg$ sample requires the photon
  to have 10 GeV or more of transverse energy, and to be in the
  central region. 
Using an event selection optimized
  for the $\ttg$ candidate sample we measure the production cross
  section of $\ttbar$ $(\sigma_{\ttbar})$, and the ratio of cross
  sections of the two samples.
   Control 
samples in the dilepton+photon  and lepton+photon+$\met$, 
channels are constructed to aid in decay product identification and
 background measurements.
We observe $\datanoflttbargcutsisojetssuper$ $\ttg$ candidate events
 compared to the standard model expectation of 
$\smnodcnoflttbargcutsisojetssuper \pm
\smdtotnoflttbargcutsisojetssuper$ events. We measure
 the $\ttg$ cross section 
($\sigma_{\ttg}$) to be
$\totxsecnoflttbargcutsisojetssuper \pm \totxsecdtotnoflttbargcutsisojetssuper$
~pb,
and the ratio of $\sigma_{\ttg}$ to $\sigma_{\ttbar}$
 to be $\ratiottgttbar \pm \dratiottgttbar$. Assuming no
$\ttg$ production, we observe a probability of 0.0015 of the background events
 alone producing $\datanoflttbargcutsisojetssuper$ events or more, 
corresponding to 
3.0 standard deviations.
\end{abstract}


\maketitle




The standard model (SM)~\cite{SM} of particle physics makes successful 
predictions of the
 production rates of physics processes that span many orders of magnitude.
  Data from $\ppbar$ collisions collected at the Tevatron have been used to
 verify many of these predictions~\cite{simona}.  
As a test of the SM,
 we measure the ratio of production cross sections of $\ttg$ to $\ttbar$.
  The ratio allows for the cancellation of systematic effects, and is a
 more sensitive test of the SM than the measurement of the production cross
 section of $\ttg$ alone.  While current data is not sufficient to 
study them in detail, the $\ttg$ coupling parameters
are sensitive to some  new physics models~\cite{UliBaur}, and will be better
measured in the future.

Top quarks are dominantly produced in pairs, with both top quarks
decaying to a $W$ boson and a $b$ quark nearly 100\% of the time.
 Their decays are classified as
dileptonic if both $W$ bosons decay to leptons, semileptonic if only
one $W$ boson decays to leptons, and hadronic if neither $W$ boson
decays to leptons. 
Selection for the $\ttg$ events in a semileptonic 
channel (including $\tau$ leptonic decays) was performed using
 $\apprluminosityfb~\invfb$ of integrated
luminosity from $\ppbar$ collisions at $\roots$ = 1.96 $\TeV$
collected using the CDF II detector~\cite{CDFII}.   In order to isolate 
non-hadronic $\ttg$ production, we require:  a high-transverse-momentum ($\pt$)
\cite{CDF_coo} lepton ($\ell$) identifed as either an electron ($e$) or a
 muon ($\mu$), 
a photon ($\gamma$), 
a $b$-tagged jet ($b$), missing transverse energy ($\met$),
large total transverse energy ($\Ht$), and three or more jets.
With these selection criteria, $\ttg$ dominates 
SM predictions~\cite{Irina_PRD}.   The total transverse energy, $\Ht$, is the
 scalar sum of the transverse energy
of electrons, muons, jets, photons, and $\met$ identified in the event.
 Furthermore, we select top-quark pair production ($\ttbar$) events by using
 nearly the same 
selection as $\ttg$, but without the photon requirement. Using similar
event selection ensures that many systematic uncertainties cancel when 
we measure the cross section ratio of $\ttg$ to $\ttbar$.

The semileptonic cross section of $\ttg$ has been measured to be
 0.15~$\pm$~0.08 pb
using data corresponding to an integrated luminosity of 
1.9$~\invfb$~\cite{Irina_PRD}.   Using the branching ratio of $W$ decays
to leptons (0.324)~\cite{PDG}
, this corresponds to a production
cross section of 0.34$~\pm$ 0.18 pb.  The cross section of $\ttbar$ production
is well-measured at 7.70$~\pm$~0.52 pb~\cite{ttbar_xsec_combined}.  However a
measurement of both $\ttbar$ and $\ttg$ cross sections with similar event
selection has not been performed.



The CDF II detector is a cylindrically symmetric magnetic
 spectrometer designed to 
study $\ppbar$ collisions at the Fermilab Tevatron.  Here we briefly
describe the components relevant for this analysis.
Tracking systems are used 
to measure the momenta of charged particles and to assist lepton 
identification. A
multi-layer system of silicon strip detectors~\cite{SVX} 
which
 identifies tracks
in the $r$ - $\phi$ and $r$ - $z$ views~\cite{CDF_coo}, and
 the central outer tracker (COT)
~\cite{COT}, are contained in a superconducting solenoid that generates
a 1.4 T magnetic field.  The COT is a 3.1 m long open-cell drift chamber
capable of making up to 96 measurements of each charged particle 
in the pseudorapidity region $|\eta| <$ 1~\cite{CDF_coo}.  Sense wires are
 arranged in 8 alternating 
axial and $\pm$ 2$\deg$ stereo superlayers with 12 wires each.  
For high-momentum tracks, the COT transverse momentum resolution is 
$\sigma_{pT}/p_T^2 \simeq$  0.0015 $\GeV^{-1}.$

Segmented calorimeters with towers arranged in a projective geometry, 
each tower consisting of an electromagnetic and a hadronic
 compartment~\cite{cal_upgrade},
cover the region $|\eta|<$ 3.6.  In this 
analysis we select photons and electrons from the central region, 
$|\eta| \lesssim 1.0$, where the central electromagnetic shower system (CES) 
makes profile measurements at shower maximum with finer 
 spatial resolution than the calorimeter. 
  Electrons are
reconstructed in the central electromagnetic calorimeter (CEM) with an
$\Et$~\cite{CDF_coo} resolution of
 $\sigma(\Et)/\Et \simeq$ 13.5$\%$/$\sqrt{\Et/\GeV} 
\oplus 1.7\%$.  Jets are identified using the hadronic and electromagnetic
calorimeter using a cone in $\eta - \phi$ space of radius $R= \sqrt{(\Delta\phi)^2 + (\Delta\eta)^2} = 0.4$.  The jet 
energy resolution is approximately $\sigma \simeq 0.1 \times \Et (\GeV) +
1~\GeV$~\cite{JetResolution} ({\it i.e.} 2.5 $\GeV$ for a 15 $\GeV$ jet).

%


Jets containing a hadron with a $b$ quark ($b$ hadrons) are identified by
 exploiting the long $b$-hadron lifetime
(c$\tau_b \approx $450 $\mu$m).   
The tracks originating from the resulting displaced vertex
 are used by the  {\sc SecVtx} ~\cite{SecVtx} algorithm to identify the 
$b$ hadron.  
The algorithm works in the region $|\eta|<$ 2, defined by the silicon system
coverage.  Jets that are identified as coming from $b$ hadrons are said to
be $b$-tagged.


Muons ($\mu$) are identified using the 
central muon (CMU), the central muon
upgrade (CMP), and the central muon extension (CMX) systems
~\cite{muon_systems}, which cover the detector region  $|\eta|<$ 1. 

Luminosity is measured using \u{C}erenkov luminosity counters in the 
range 3.7 $ <|\eta| <$ 4.7.  The uncertainty in the luminosity has been
estimated to be 6$\%$, where 4.4$\%$ comes from the acceptance and 
operation of the luminosity monitor, and 4$\%$ comes from the uncertainty
on the inelastic cross section of $\ppbar$~\cite{CLC}.

A three-level online event selection system (trigger)~\cite{CDFII} selects
 events with a high-$\pt$ lepton in the central region. 
The trigger system selects electron candidates from 
clusters of energy in the central electromagnetic calorimeter. Electrons 
are distinguished from photons by requiring a COT track associated with the 
clusters. The muon trigger requires a COT track that extrapolates to a 
track segment (``stub'') in the muon detectors.




A muon candidate passing 
our 
 selection criteria must have:
 a well-measured 
track in the COT, energy deposited in the calorimeter consistent with 
minimum-ionization expectations, a muon stub in 
both the CMU and CMP, or in the CMX, 
consistent with the extrapolated COT track, and COT timing consistent 
with a track from a $\ppbar$ collision. 

An electron candidate passing 
 our selection criteria must have: a high-quality track with $\pt>0.5~\Et$, 
unless $\Et > 100$ $\GeV$, in which case the $\pt$ threshold is set to 25 
$\GeV$, a good transverse shower profile that matches the extrapolated 
track position, a lateral sharing of energy in the two calorimeter 
towers containing the electron shower consistent with that expected for
an electromagnetic (EM) shower, and 
 minimal leakage into the hadron calorimeter~\cite{hadoem}.

Photon candidates are required to have $\Etgamma>$10$~\GeV$,
 no track with $\pt>1$ $\GeV$ 
and at most one track with $\pt<1$ $\GeV$, pointing at the EM cluster,
 good profiles in both transverse dimensions at shower
maximum, and minimal leakage into the hadron
calorimeter~\cite{hadoem}.  The detected tracks have a minimum
 $\pt$ of 0.35 $\GeV$ due to the magnetic field curling up lower $\pt$
particles.
The photons are only reconstructed in the CEM and have $|\eta| <$ 1.0.

To reduce background from photons or leptons that originate from decays of
hadrons produced in jets, both the photon and the lepton in each event
are required to be ``isolated''. The $\Et$ deposited in the
calorimeter towers in a cone in $\eta-\varphi$ space~\cite{CDF_coo} of
 $R = 0.4$ around the photon or lepton position is summed, and the
$\Et$ due to the photon or lepton is subtracted. The remaining $\Et$
is required to be less than $2.0~\GeV+0.02\times($$\Et$ - 20$~\GeV)$ for a
photon, or less than 10\% of the $\Et$ for electrons or $\pt$ for
muons. In addition, for photons, the sum of the $\pt$ of all tracks in
the cone must be less than $2.0~\GeV+0.005\times$$\Et$.


Missing transverse energy, $\met$, is calculated from the observed
 calorimeter-tower
 energies in the region $|\eta| < 3.6$ ~\cite{CDF_met}. Corrections are
 then made
to the $\met$ for non-uniform calorimeter response~\cite{jet_corr} for
jets with uncorrected $\Et > 15$ $\GeV$ and $\eta < 2$, and for
muons with $\pt > 20$ $\GeV$.  

Events for the analysis are selected by requiring
 a central $e$ or $\mu$ with 
$\Etlepton>20$ $\GeV$ originating less than 60 cm along the beam-line from 
the detector center and passing the 
  criteria listed above.  We 
further require events to have at least one of the following objects: 
a jet with $\Etjet>$ 15 $\GeV$,
$\met >$ 20 $\GeV$, an additional lepton, or a central $\gamma$
 with $\Et>$ 10 $\GeV$.


The first measurement we perform is in the $\ttbar$ signal sample, which
 requires
an event to contain:
 $\met>20~\GeV$, 
a 
 lepton, 
a $b$-tagged jet, 
$\Ht> 200~\GeV$, $\Njets \geq$ 3~\cite{top_dileptonic} 
 (including the $b$-tagged jet) , and
transverse mass of the lepton and $\met$  to be greater than 20 $\GeV$
for the electron channel, and 10 $\GeV$ for the muon channel. 
Transverse mass for the $\met$
 and lepton is defined as:
$\sqrt{2({E}_{\ell,T}\times\met - { E}_{\ell,x}\times\met_x -{E}_{\ell,y}\times\met_y)}$.
The selection criteria is inclusive, so if an event contains an
additional lepton or a photon it is also accepted as a signal event.
The highest-$\pt$ lepton determines if the event is an electron or
muon event.


Events in the $\ttg$ signal sample are selected by requiring
$\met > 20~\GeV$, a 
 lepton, a $b$-tagged jet,
$\Ht > 200~\GeV$, $\Njets \geq$ 3 (including the $b$-tagged jet.), and a photon with $\Et>10~\GeV$. 
For all photons we require the $\chi^2$ of the CES shower profile be less than
20. To further suppress backgrounds,
 photons with $\Et$ between 10 and 25 $\GeV$ must have a 
$\chi^2$ of the CES shower profile less than
6; we discuss how $\chi^2$ is calculated
below.  Similar to the $\ttbar$ analysis, the selection is inclusive. 
The selection
criteria are identical to the previous $\ttg$ cross section measurement
~\cite{Irina_PRD}, with the exception of the low-$\Et$ photon $\chi^2$ 
requirements.

The primary difference between the $\ttbar$ and $\ttg$ selection, other than
 the photon selection, is the requirement of
a transverse mass selection for $\ttbar$.  In the $\ttbar$ selection, the 
low-transverse-mass region is not well-modeled with background estimation
methods.  The $\ttg$ sample does not suffer from this deficiency, and we
do not use the transverse mass selection criterion to keep acceptance of
 signal events high, this behavior is also seen in the $\lgmet$ control sample 
described below.

Control samples are identified by selecting events with a 
 lepton, a 
 photon, and $\met > 20~\GeV$ ($\lgmet$), or two oppositely charged 
same-flavor central leptons, a 
  photon, and a three-body mass
consistent with the $Z$ boson ($\llg$).    These control samples are used
to define the above CES $\chi^2$ selection for photons.

The  $\chi^2$ value of photons is based on the lateral shower shapes
observed in the CES compared to that predicted from a sample of test-beam
electrons.  Using
the control samples we identify an additional selection criterion on 
photons~\cite{bensThesis}.
  It should be noted
that the $\ttg$ sample
 contains $\datanoflttbargcutsisojetssuper$ events and is a subset
 of the $\datanoflwgiso$ events in the
 $\lgmet$ sample. The $\llg$ sample contains 1344 events.
While the samples are not independent, optimizing photon identification 
selection criteria using the $\lgmet$
sample should be minimally affected by the presence of $\ttg$ events.

The dominant SM sources of events with a lepton, photon, and significant
 $\met$, 
not including particle misidentifications,  are $\ttg$
production and $\Wg$+heavy flavor (HF), in which a $W$ boson decays
leptonically ($\ell \nu$) and a photon is radiated from an
initial-state or final-state quark, the $W$ boson, or a charged final-state
lepton~\cite{CDF_WZgamma}.  In this paper, HF includes: $\ccbar$, $\bbbar$, 
and $c$.
Similarly, for events in the $\ttbar$
selection, the dominant source of events is due to $\ttbar$ production
and $W$+HF production.

The production of  $\ttg$ events with semileptonic and dileptonic
decays, as well as the SM background of single-top events
and associated production
of a $\Wg$+HF
is estimated from leading-order (LO) matrix-element  Monte Carlo
simulations 
(MC) event generator 
{\sc MadGraph}~\cite{MadGraph}. 
Events for all production and decays of $\ttbar$, $WW, WZ,$ and $ZZ$
 signals are generated with {\sc pythia}~\cite{Pythia}.  The production of
 $W$+HF, 
as well as
$Z + \bbbar$, and $Z\rrr\tau\tau$ decays are generated with 
{\sc alpgen}~\cite{Alpgen}.  Then the
 events are processed with the same reconstruction
 and analysis
codes used for the data.    Backgrounds from
$ZZ$ are estimated to be negligible to the $\ttg$ signature.


Initial state radiation is simulated by the {\sc pythia} shower Monte
Carlo simulation code tuned so as to reproduce the underlying
event~\cite{PythiaTuneA}. All of the generated samples are then passed through a full
simulation of the detector, 
 then reconstructed
with the same reconstruction code used for the data.

\begin{figure}[!b]
\begin{center}
\hspace*{-0.1in}
\includegraphics*[angle=90,width=0.50\textwidth]{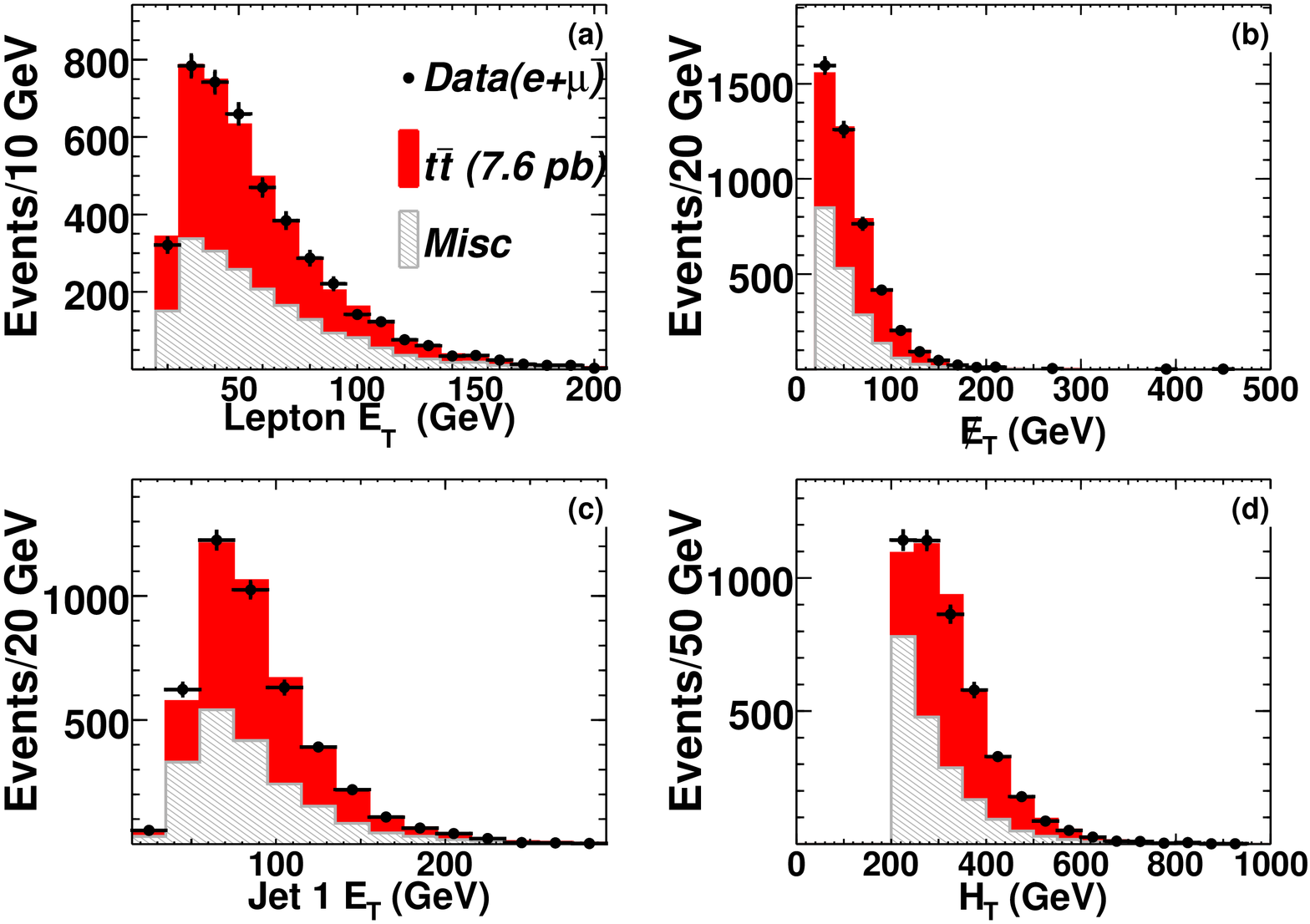}
\end{center}
\vskip-1cm
\caption{ The distributions for events in the $\ttbar$ sample (points)
of 
  a) the $\Et$ of the lepton; 
  b) the missing transverse energy $\met$;
  c) the $\Et$ of the highest $\Et$ jet; 
  and
  d) the total transverse energy $\Ht$.
  The histograms show the expected SM contributions from top production 
($\ttbar$), and miscellaneous backgrounds (Misc), which include: 
diboson production, single top, $W$+HF and $Z$+HF production as
 well as
 jets misidentified as leptons (QCD), and misidentified $b$ tags.}
\label{figure1ttbar}
\end{figure}

\begin{figure}[!b]
\begin{center}
\hspace*{-0.1in}
\includegraphics*[angle =90,width=0.50\textwidth]{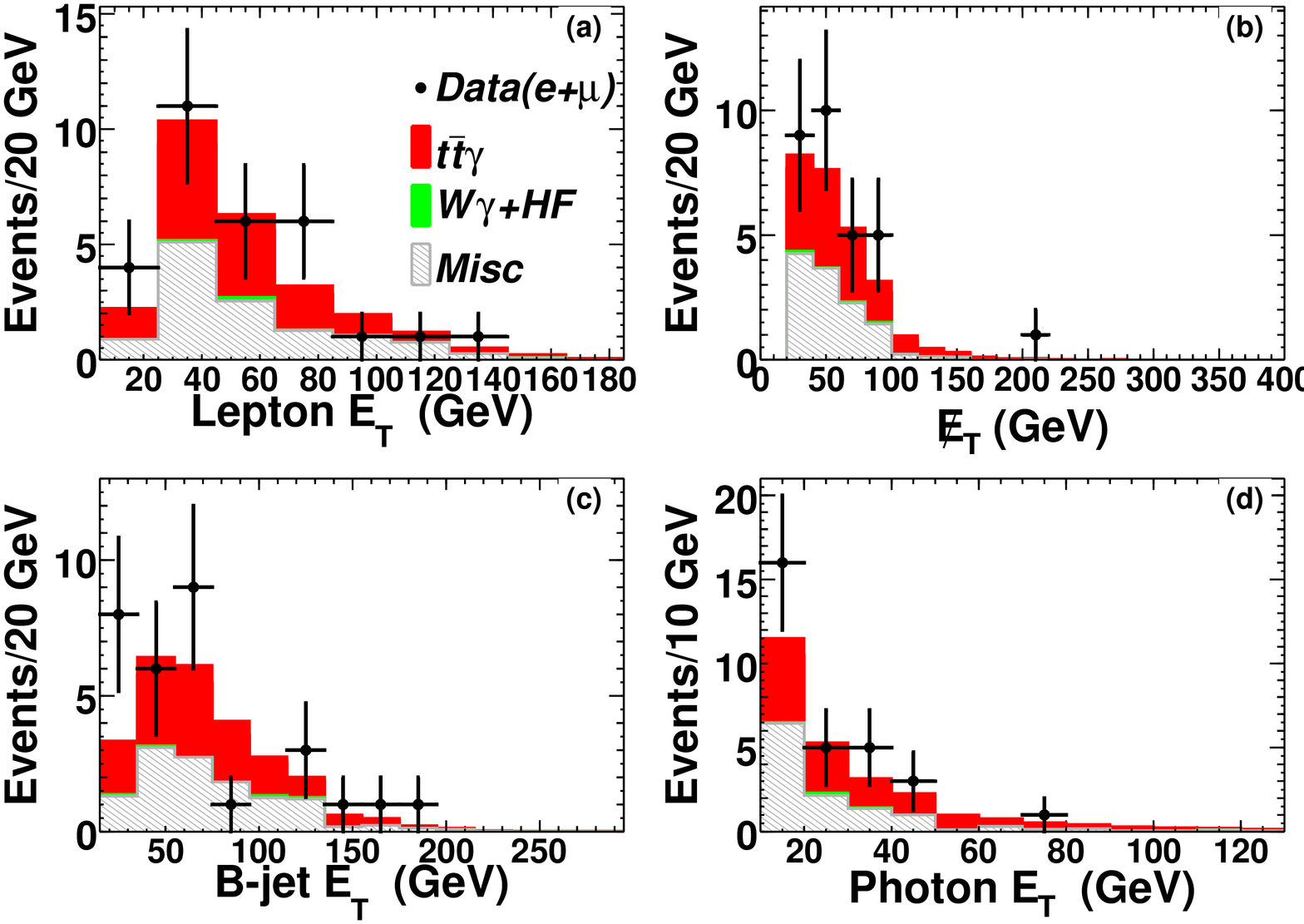}
\end{center}
\vskip-1cm
\caption{ The distributions for events in the $\ttg$ sample (points)
  in
  a) the $\Et$ of the lepton; 
  b) the  missing transverse energy, $\met$; 
  c) the $\Et$ of the $b$ jet;
  and
  d)  the $\Et$ of the photon.
  The histograms show the expected SM contributions from radiative 
top production 
($\ttg$), $W\gamma$ production with heavy flavor (HF), and miscellaneous backgrounds (Misc), which include: 
$WW$ and $WZ$ production as well as
jets, $\tau$ leptons, electrons, and jets misidentified as photons,
 jets misidentified as leptons (QCD), and misidentified $b$ tags.}
\label{figure3}
\end{figure}

\begin{figure}[t]
\begin{center}
\hspace*{-0.1in}
\includegraphics*[angle=90,width=0.50\textwidth]{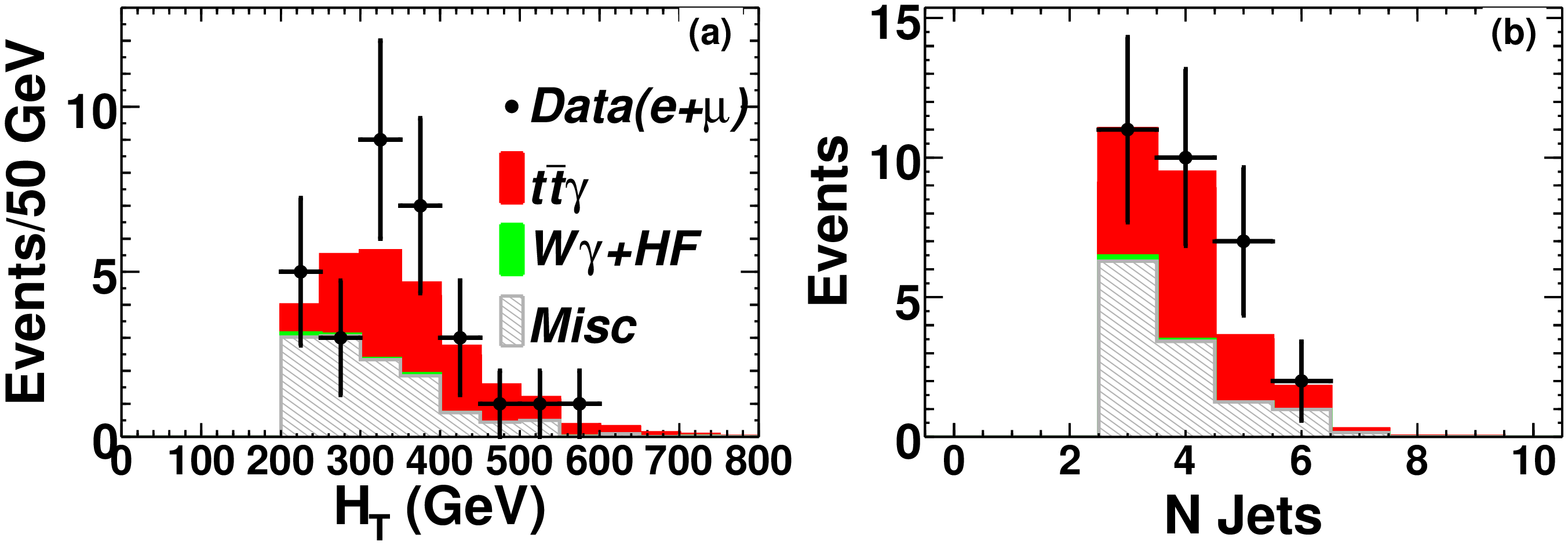}
\end{center}
\vskip-1cm
\caption
{The distributions for events in the $\ttg$ sample (points) of a)
the total transverse energy $\Ht$, the
sum of the transverse energies of the lepton, photon, jets and $\met$,
for the $\ttg$ events; b)  the total number of jets .
   The histograms show the expected SM contributions from radiative top-quark
pair production
($\ttg$), $W\gamma$ production with heavy flavor ($\Wg$+HF), and miscellaneous backgrounds (Misc), which include: 
SM $WW$ and $WZ$ production as well as
jets, $\tau$ leptons, electrons, and jets misidentified as photons,
 jets misidentified as leptons (QCD), and misidentified $b$ tags.}
\label{figure4}
\end{figure}

The expected contributions from $\ttbar$, $W+HF$, single-top, $Z+\bbbar$,
and $Z\rrr\tau\tau$ production to the $\ttbar$ search 
 are given in Table~\ref{table0}, and
the expected contributions from $\ttg$ and $\Wg+HF$ production to the
 $\ttg$ search are given in Table~\ref{table2}. Additional contributions
from misidentification backgrounds, described below, are also shown in the
Tables.
 Figure~\ref{figure1ttbar} shows 
kinematic distributions for the $\ttbar$ sample, and 
 Figs.~\ref{figure3} and \ref{figure4} show distributions for
 events in the $\ttg$ sample. There is good agreement
between data and standard-model predictions. We show the data 
and background predictions combined for both electron and muon events; there
is good agreement in both channels individually, as shown 
in~\cite{bensThesis}.

\begin{table}[!t]

 \caption {Summary for predicted and observed 
events in the $\ttbar$ ($\lmetb$ + $\Ht >$ 200 $\GeV$
+ $\Njets \geq 3$) signal sample. }

  \begin{center}

\begin{tabular}{lccc}
\hline
\multicolumn{4}{c} {
Predicted and Observed 
$\ttbar$ Events }\\
\hline
SM Source & $e b\met$ & $\mu b\met$ & $(e+\mu) b\met$ \\
\hline
$\ttbar $                                        & 

$\ttnofettbarcutsisojets  \pm  \totttdsysnofettbarcutsisojets   $   &
$\ttnofmuttbarcutsisojets  \pm \totttdsysnofmuttbarcutsisojets  $   &
$\ttnoflttbarcutsisojets \pm \totttdsysnoflttbarcutsisojets   $   \\

$WW$ &

$\wwnofettbarcutsisojets  \pm \totwwdsysnofettbarcutsisojets  $     &
$\wwnofmuttbarcutsisojets \pm \totwwdsysnofmuttbarcutsisojets $    &
$\wwnoflttbarcutsisojets  \pm \totwwdsysnoflttbarcutsisojets  $     \\

$WZ$ &
$\wznewnofettbarcutsisojets  \pm \totwznewdsysnofettbarcutsisojets  $     &
$\wznewnofmuttbarcutsisojets  \pm \totwznewdsysnofmuttbarcutsisojets  $     &
$\wznewnoflttbarcutsisojets  \pm \totwznewdsysnoflttbarcutsisojets  $     \\

$ZZ$ &

$\zznofettbarcutsisojets  \pm \totzzdsysnofettbarcutsisojets  $     &
$\zznofmuttbarcutsisojets \pm \totzzdsysnofmuttbarcutsisojets $    &
$\zznoflttbarcutsisojets  \pm \totzzdsysnoflttbarcutsisojets  $     \\

$W bb$  &

$\wbbincnofettbarcutsisojets  \pm \wbbincdtotnofettbarcutsisojets  $     &
$\wbbincnofmuttbarcutsisojets \pm \wbbincdtotnofmuttbarcutsisojets $    &
$\wbbincnoflttbarcutsisojets  \pm \wbbincdtotnoflttbarcutsisojets  $     \\

$W cc$   &
$\wccincnofettbarcutsisojets  \pm \wccincdtotnofettbarcutsisojets  $     &
$\wccincnofmuttbarcutsisojets \pm \wccincdtotnofmuttbarcutsisojets $    &
$\wccincnoflttbarcutsisojets  \pm \wccincdtotnoflttbarcutsisojets  $     \\


$W c$   &

$\wcincnofettbarcutsisojets  \pm  \wcincdtotnofettbarcutsisojets $     &
$\wcincnofmuttbarcutsisojets \pm  \wcincdtotnofmuttbarcutsisojets$    &
$\wcincnoflttbarcutsisojets  \pm \wcincdtotnoflttbarcutsisojets  $     \\

Single top (s-ch.)   &
$\stopsnofettbarcutsisojets   \pm \totstopsdsysnofettbarcutsisojets  $     &
$\stopsnofmuttbarcutsisojets  \pm \totstopsdsysnofmuttbarcutsisojets $    &

$\stopsnoflttbarcutsisojets \pm \totstopsdsysnoflttbarcutsisojets  $ \\



Single top (t-ch.)  &

$\stoptnofettbarcutsisojets \pm \totstoptdsysnofettbarcutsisojets  $     &
$\stoptnofmuttbarcutsisojets \pm \totstoptdsysnofmuttbarcutsisojets  $     &
$\stoptnoflttbarcutsisojets \pm \totstoptdsysnoflttbarcutsisojets  $     \\

$(Z \rrr \ell\ell) b\bar{b}$   &

$\zbbincnofettbarcutsisojets \pm \zbbincdtotnofettbarcutsisojets $ &
$\zbbincnofmuttbarcutsisojets \pm \zbbincdtotnofmuttbarcutsisojets $ &
$\zbbincnoflttbarcutsisojets \pm \zbbincdtotnoflttbarcutsisojets $ \\

$Z \rrr \tau\tau$   &


$\zttincnofettbarcutsisojets \pm \zttincdtotnofettbarcutsisojets $  &
$\zttincnofmuttbarcutsisojets \pm \zttincdtotnofmuttbarcutsisojets $  &
$\zttincnoflttbarcutsisojets \pm \zttincdtotnoflttbarcutsisojets $  \\

Mistags &

$\mistagnofettbarcutsisojets  \pm \totmistagdsysnofettbarcutsisojets     $     &
$\mistagnofmuttbarcutsisojets \pm \totmistagdsysnofmuttbarcutsisojets    $    &
$\mistagnoflttbarcutsisojets  \pm \totmistagdsysnoflttbarcutsisojets  $     \\

QCD &

$\jqcdnofettbarcutsisojets  \pm \totjqcddsysnofettbarcutsisojets    $     &
$\jqcdnofmuttbarcutsisojets \pm \totjqcddsysnofmuttbarcutsisojets    $    &
$\jqcdnoflttbarcutsisojets  \pm \totjqcddsysnoflttbarcutsisojets  $     \\


\hline
Total Predicted                    &

$ \smnodcnofettbarcutsisojets  \pm \smdtotnofettbarcutsisojets $    &
$ \smnodcnofmuttbarcutsisojets  \pm \smdtotnofmuttbarcutsisojets $    &
$ \smnodcnoflttbarcutsisojets  \pm \smdtotnoflttbarcutsisojets $    \\


\hline\hline
Observed                       & 
$\datanofettbarcutsisojets$                          & 
$\datanofmuttbarcutsisojets$                         & 
$\datanoflttbarcutsisojets$                          \\
\hline
\end{tabular}

  \end{center}

\label{table0}
\end{table}

\begin{table}[!b]
\caption
{Summary for $\ttg$ $(\lgmet b + H_T>200~\GeV + N_{jets} \geq
3$). Monte Carlo samples listed in the table are given as they were generated
(\textit{e.g.}, $WW$ was not generated with an associated photon.)
 Backgrounds from 
$ZZ$
are found to be negligible.  }
\begin{center}
\begin{tabular}{lccc}
\hline
\multicolumn{4}{c} {
Predicted and Observed $\ttg$ Candidate Events}\\
\hline
SM Source & $e\gamma b\met$ & $\mu\gamma b\met$ & $(e+\mu)\gamma b\met$ \\
\hline
$\ttg (semilep)$                                        & 

$\ttasemnofettbargcutsisojetssuper  \pm \totttasemdsysnofettbargcutsisojetssuper$   &
$\ttasemnofmuttbargcutsisojetssuper \pm \totttasemdsysnofmuttbargcutsisojetssuper$  &
$\ttasemnoflttbargcutsisojetssuper  \pm \totttasemdsysnoflttbargcutsisojetssuper$   \\

$\ttg (dilep.)$                                        & 

$\ttadilnofettbargcutsisojetssuper  \pm \totttadildsysnofettbargcutsisojetssuper  $   &
$\ttadilnofmuttbargcutsisojetssuper \pm \totttadildsysnofmuttbargcutsisojetssuper $  &
$\ttadilnoflttbargcutsisojetssuper  \pm \totttadildsysnoflttbargcutsisojetssuper  $   \\

$W c\gamma$                         & 

$\wcanofettbargcutsisojetssuper  \pmasym{\totwcadsysnofettbargcutsisojetssuper}{0}  $     &
$\wcanofmuttbargcutsisojetssuper \pmasym {\totwcadsysnofmuttbargcutsisojetssuper}{0} $    &
$\wcanoflttbargcutsisojetssuper  \pmasym {\totwcadsysnoflttbargcutsisojetssuper}{0}  $     \\

$W cc\gamma$                         & 

$\wccanofettbargcutsisojetssuper  \pmasym {\totwccadsysnofettbargcutsisojetssuper}{0}  $     &
$\wccanofmuttbargcutsisojetssuper \pm \totwccadsysnofmuttbargcutsisojetssuper $    &
$\wccanoflttbargcutsisojetssuper  \pm \totwccadsysnoflttbargcutsisojetssuper  $     \\

$W bb\gamma$                         & 

$\wbbanofettbargcutsisojetssuper  \pm \totwbbadsysnofettbargcutsisojetssuper  $     &
$\wbbanofmuttbargcutsisojetssuper \pm \totwbbadsysnofmuttbargcutsisojetssuper $    &
$\wbbanoflttbargcutsisojetssuper  \pm \totwbbadsysnoflttbargcutsisojetssuper  $     \\

$WZ$                                          & 

$\wznofettbargcutsisojetssuper  \pm \totwzdsysnofettbargcutsisojetssuper  $     &
$\wznofmuttbargcutsisojetssuper \pm \totwzdsysnofmuttbargcutsisojetssuper $    &
$\wznoflttbargcutsisojetssuper  \pm \totwzdsysnoflttbargcutsisojetssuper  $     \\

$WW$                                          & 

$\wwnofettbargcutsisojetssuper  \pm \totwwdsysnofettbargcutsisojetssuper  $     &
$\wwnofmuttbargcutsisojetssuper \pm \totwwdsysnofmuttbargcutsisojetssuper $    &
$\wwnoflttbargcutsisojetssuper  \pm \totwwdsysnoflttbargcutsisojetssuper  $     \\

Single Top (s-chan)                                          & 

$\stopsnofettbargcutsisojetssuper  \pm \totstopsdsysnofettbargcutsisojetssuper  $     &
$\stopsnofmuttbargcutsisojetssuper \pm \totstopsdsysnofmuttbargcutsisojetssuper $    &
$\stopsnoflttbargcutsisojetssuper  \pm \totstopsdsysnoflttbargcutsisojetssuper  $     \\

Single Top (t-chan)                                          & 

$\stoptnofettbargcutsisojetssuper  \pm \totstoptdsysnofettbargcutsisojetssuper  $     &
$\stoptnofmuttbargcutsisojetssuper \pm \totstoptdsysnofmuttbargcutsisojetssuper $    &
$\stoptnoflttbargcutsisojetssuper  \pm \totstoptdsysnoflttbargcutsisojetssuper  $     \\

$\tau \rrr \gamma$ fake                                        & 

$\taufakeenofettbargcutsisojetssuper  \pm \tottaufakeedsysnofettbargcutsisojetssuper  $   &
$\taufakeenofmuttbargcutsisojetssuper \pm \tottaufakeedsysnofmuttbargcutsisojetssuper $  &
$\taufakeenoflttbargcutsisojetssuper  \pm \tottaufakeedsysnoflttbargcutsisojetssuper  $   \\


Jet faking $\gamma$ & 

$\corrjfakepnofettbargcutsisojetssuper  \pm \totjfakepdsysnofettbargcutsisojetssuper$       &
$\corrjfakepnofmuttbargcutsisojetssuper \pm \totjfakepdsysnofmuttbargcutsisojetssuper$      &
$\corrjfakepnoflttbargcutsisojetssuper  \pm \totjfakepdsysnoflttbargcutsisojetssuper$       \\


Mistags                                       & 
$\corrmistagnofettbargcutsisojetssuper  \pm \totmistagdsysnofettbargcutsisojetssuper$       &
$\corrmistagnofmuttbargcutsisojetssuper \pm \totmistagdsysnofmuttbargcutsisojetssuper$      &
$\corrmistagnoflttbargcutsisojetssuper  \pm \totmistagdsysnoflttbargcutsisojetssuper$       \\

QCD     & 

$\corrjqcdnofettbargcutsisojetssuper  \pm \totjqcddsysnofettbargcutsisojetssuper $   &
$\corrjqcdnofmuttbargcutsisojetssuper \pm \totjqcddsysnofmuttbargcutsisojetssuper$   &
$\corrjqcdnoflttbargcutsisojetssuper  \pm \totjqcddsysnoflttbargcutsisojetssuper $   \\


$\eemet b$, $e\goes\gamma$                      & 
$\correfpnofettbargcutsisojetssuper \pm \totefpdsysnofettbargcutsisojetssuper$    &
--                                       &
$\correfpnofettbargcutsisojetssuper \pm \totefpdsysnofettbargcutsisojetssuper$    \\
$\mu e\met b$, $e\goes\gamma$                      & 
--                                       &
$\correfpnofmuttbargcutsisojetssuper \pm \totefpdsysnofmuttbargcutsisojetssuper$    &
$\correfpnofmuttbargcutsisojetssuper \pm \totefpdsysnofmuttbargcutsisojetssuper$    \\


\hline
Total  Predicted                    & 

$ \smnodcnofettbargcutsisojetssuper  \pm \smdtotnofettbargcutsisojetssuper $    &
$ \smnodcnofmuttbargcutsisojetssuper  \pm \smdtotnofmuttbargcutsisojetssuper $    &
$ \smnodcnoflttbargcutsisojetssuper  \pm \smdtotnoflttbargcutsisojetssuper $    \\



\hline\hline
Observed                        & 
$\datanofettbargcutsisojetssuper$                          & 
$\datanofmuttbargcutsisojetssuper$                         & 
$\datanoflttbargcutsisojetssuper$                          \\
\hline
\end{tabular}

\end{center}

\label{table2}
\end{table}

High $\pt$ photons are copiously produced in hadron jets
initiated by a scattered quark or gluon.  
In the $\ttg$ sample, the number of events in which a jet is misidentified 
as a photon (jet faking photon)  is estimated by removing the isolation 
requirements on the photon.  We fit the resulting isolation distribution  using
 signal  and background templates obtained from data.  The signal template
is constructed  using electrons from $\Zgstar \rrr ee$ events, and a
 background template
is made from a  QCD enriched sample~\cite{Andrei_PRL,bensThesis}.  

The expected number of events in which an electron is misidentified as
a photon in the $\ttg$ signature is determined by measuring the
electron $\Et$ spectrum in the $\lmetb$+$e$+large $\Ht$, and $\geq$ 3
jets sample (events of this type are created from dileptonic, $ee$ or $e\mu$,
 $\ttbar$ and diboson decays), and then multiplying by the
 probability that an electron is misidentified as a photon.
  The latter is measured in data using
$\Zgstar \rightarrow ee$ events that are misreconstructed as
$\Zgstar\rrr e\gamma$.



To estimate the number of $b$-tagged jets that are in reality mistagged
 light-quark jets (mistags), each jet in the $\lgmet + \geq$ 3 jets
 and high-$\Ht$ sample is 
weighted by a mistag rate.  The mistag rate per jet is
 measured using a large inclusive-jet sample.  For the $\ttbar$ sample, a
 similar procedure is used.  Each jet in the signal samples has a
 corresponding  probability to be identified as a $b$-tagged jet.  
In all cases,
 however, the resulting  prediction is  overestimated because we count as
 mistags, events which have true heavy flavor jets  
 (\textit{i.e.} events due to $\ttbar$ events may
be mistagged, but they will be accounted for in the MC). The fraction's 
denominator is computed by finding the total number 
of $\lgmet$ $\geq$ 3 jets (or $\ttbar$ analogue) events.  Its 
numerator is the difference between the denominator and the number of 
events in the sample with $b$-tagged jets predicted by MC simulations.  
    The
 fraction is the amount of
 events 
 that have
no true heavy flavor content relative to the size of $\lgmet$ $\geq$ 3 jets
(or $\ttbar$ analogue) sample; it is used to scale our mistag
 estimate.  This scaled background estimate removes events which contain
 actual heavy flavor content from the mistag total; the scale factor is 
nearly 60\%.



The background due to events in which a jet is misidentified as a lepton (QCD)
 is estimated
using ``non-electrons''.  ``Non-electrons'' are jets which are
 kinematically similar
to electrons, but which fail a pair of selection criteria normally passed by 
electrons (regardless of energy) 
 such as: EM shower-shape, energy over momentum, or 
isolation.  
``Non-electrons''
play the role of leptons in our model of the QCD background.
  The remaining object selection
is unchanged. 
 In each  
signature, a template of signal events in data with $\met <$ 20 $\GeV$
are  fit
to the sum of MC backgrounds and a scaled ``non-electron'' signal. 
 The QCD background is the
 sum of the scaled ``non-electron'' events in the $\met >$ 20 $\GeV$ region 
expected from the fit.

To avoid double counting, the total background yields are corrected by removing
 the predicted number of events with two objects misidentified.
  Each of the aforementioned data-driven
background estimates accounts for a background process where one object in
the event is misidentified.   Events with two  misidentified objects
would be counted in a pair of background estimates.  
In the $\ttbar$ sample,  double counting is accounted for by removing the
 QCD background from the mistagging background, and vice versa.


The background from tau leptons, which decay to hadrons, which decay to 
photons is a background
estimated from the $\ttbar$ MC 
 sample by selecting
$\tau\rrr~hadrons~\rrr\gamma$ events using MC information.



The $\ttbar$ event detection efficiency and acceptance are calculated
using the MC simulation which has 
 all decays of  $\ttbar$.  The uncertainty on the $\ttbar$
 cross section is dominated by systematic uncertainties.  The $\ttg$
event detection efficiency and acceptance are calculated using both 
semileptonic, and
dileptonic decays of the pair of top quarks in the decay.  The total $\ttg$
cross section is then calculated assuming that $\ttg$ has the
 same branching ratio to semileptonic and
dileptonic decays as $\ttbar$ pair production.  The uncertainty 
in the $\ttg$ cross section measurement is dominated by statistics.

Systematic uncertainties have been calculated by varying 
detector efficiencies and resolutions within known uncertainties and 
evaluating the change in our measurements.  These uncertainties are added
in quadrature when independent, and summed when positively (or negatively)
 correlated.  
The largest
 uncertainties, given in descending order, are due to luminosity,
 $b$-hadron tagging efficiencies and, for the $\ttg$ sample, 
photon identification.

We observe  $\datanoflttbargcutsisojetssuper$ $\ttg$ candidate
 events compared to an expectation of $\smnodcnoflttbargcutsisojetssuper \pm
\smdtotnoflttbargcutsisojetssuper$.   We observe $\datanoflttbarcutsisojets$
 $\ttbar$ events, with an expectation of 
$\smnoflttbarcutsisojets \pm \smdtotnoflttbarcutsisojets$.
Assuming the difference between the non-$\ttbar$ background estimate
and the number of observed events is due to SM $\ttbar$ production,
we measure the $\ttbar$ cross section to be 
$\xsecnoflttbarcutsisojets~ \pm~ 
\xsecdnoflttbarcutsisojets {\rm~(stat)~} \pm~
\xsecdsysnoflttbarcutsisojets  {\rm~(sys)~} \pm~
\xsecdlumnoflttbarcutsisojets  {\rm~(lum)~} $pb.
The theoretical production cross section of $\ttbar$ at the Tevatron is
 $7.08 \pmasym{0.00}{0.32}~\pmasym{0.36}{0.27}$ pb~\cite{ttbartheory}.  The 
first uncertainty comes from scale uncertainty around $\mu$ = m$_{top}$, and
the second is due to parton distribution function uncertainties.

If one assumes that $\ttg$ is not allowed in the SM, and there are no 
new physics processes contributing to this sample, the probability that
the background events alone will produce $\datanoflttbargcutsisojetssuper$ 
or more events is 0.0015 (3.0 standard deviations).  This is the first
experimental evidence for $\ttg$ production.  Assuming the difference
 between the background estimate 
and the number of observed events is due to SM $\ttg$ production,
we  measure 
the $\ttg$ cross section to be 
$\totxsecnoflttbargcutsisojetssuper \pm
\totxsecdnoflttbargcutsisojetssuper  {\rm~(stat)~} \pm
 \totxsecdsysnoflttbargcutsisojetssuper {\rm~(sys)~} \pm
 \totxsecdlumnoflttbargcutsisojetssuper {\rm~(lum)~} $pb.  
The $\ttg$ event detection efficiency and acceptance are calculated
using the MC sample requiring at least one $W$ boson decaying leptonically. 
The acceptance times efficiency, using both semileptonic and dileptonic modes,
 for this $\ttg$ signal is 0.015 $\pm$ 0.002. The uncertainty on the measured
 cross section
is dominated by the statistical uncertainties associated with the
small number of events observed.
A theoretical value for the non-hadronic decays of $\ttg$
 (sum of all three lepton flavors)
cross section $\sigma_{\ttg} = 0.071 \pm 0.011~$pb is
obtained from leading order (LO) {\sc MadGraph} semileptonic cross 
section $\sigma_{\ttg} =
0.0726$~pb multiplied by a K-factor to find the next to leading order 
(NLO); K-factor $=\sigma_{NLO}/\sigma_{LO}=0.977$
~\cite{ttgKfactor}.  The next to leading order theoretical 
total cross section for $\ttg$ is thus
 $\sigma^{total}_{\ttg}$ = 0.17 $\pm$ 0.03 pb.  


The ratio between the production cross sections of $\ttg$ 
and $\ttbar$ is measured to be $\Re = \ratiottgttbar \pm \dratiottgttbar$ 
which agrees with the SM prediction of
$\Re = $ 0.024 $\pm {0.005}$, obtained from theoretical predictions of 
the cross sections of $\ttg$ and $\ttbar$.  When measuring $\Re$ many of the
systematic uncertainties nearly cancel, such as those due to: 
lepton identification,
$b$ hadron identification, jet energy scale, and luminosity
 uncertainties. However, 
other systematic uncertainties do not cancel out completely
 such as: QCD systematic uncertainties, and
 photon identification and acceptance uncertainties.  
The total systematic uncertainties combine to less than 10\% however the 
statistical uncertainty is the dominant contribution to the total uncertainty.


In conclusion, we have performed  a search for $\ttg$, which is the
dominant standard model process that produces the event signature of
lepton + photon + $\met$ + $b$-jets with
 large total transverse energy and $\Njets~\geq $ 3. We
 find that the numbers of events observed are consistent 
with SM predictions. We obtain a $\ttg$
cross section 
$\sigma_{\ttg}$ = $\totxsecnoflttbargcutsisojetssuper \pm \totxsecdtotnoflttbargcutsisojetssuper$ pb, and 
the ratio of production
cross sections of $\ttg$ to $\ttbar$ 
= $\ratiottgttbar \pm
 \dratiottgttbar$.


\begin{acknowledgments}
We thank the Fermilab staff and the technical staffs of the participating institutions for their vital contributions.  Uli Baur,
Frank Petriello, Alexander Belyaev, Edward Boos, Lev Dudko, Tim
Stelzer, and Steve Mrenna were extraordinarily helpful with the SM
predictions.   
This work was supported by the U.S. Department of Energy and National Science Foundation; the Italian Istituto Nazionale di Fisica Nucleare; the Ministry of Education, Culture, Sports, Science and Technology of Japan; the Natural Sciences and Engineering Research Council of Canada; the National Science Council of the Republic of China; the Swiss National Science Foundation; the A.P. Sloan Foundation; the Bundesministerium f\"ur Bildung und Forschung, Germany; the Korean World Class University Program, the National Research Foundation of Korea; the Science and Technology Facilities Council and the Royal Society, UK; the Institut National de Physique Nucleaire et Physique des Particules/CNRS; the Russian Foundation for Basic Research; the Ministerio de Ciencia e Innovaci\'{o}n, and Programa Consolider-Ingenio 2010, Spain; the Slovak R\&D Agency; the Academy of Finland; and the Australian Research Council (ARC). 
\end{acknowledgments}


\begin{thebibliography}{99}
\bibitem{CDF_coo} The CDF coordinate system of 
$r$, $\varphi$, and $z$ is cylindrical, with the $z$-axis along the
proton beam. The pseudorapidity is $\eta = -\ln(\tan(\theta/2))$.
 Transverse momentum and energy are defined as $\pt =
p\sin\theta$ and $\Et = E\sin\theta$, respectively.  
We use the convention that ``momentum'' refers to $pc$ 
and ``mass'' to $mc^2$.  The coordinate $y$ points up and down, and
$x$ completes the right-handed system.
\bibitem{hadoem} 
The fraction of electromagnetic energy  allowed to leak into
 the hadron compartment $\rm E_{had}/E_{em}$ must be less than $\rm
0.055\plus0.00045\times E_{em}(\GeV)$ for central electrons, less than
0.05 for electrons in the end-plug calorimeters, less than max[0.125,
$\rm 0.055\plus0.00045\times E_{em}(\GeV)$] for photons.
\bibitem{CDF_met}
%
Missing $\rm E_T$ ($\metvec$) is defined by $\metvec = -\sum_{i} E_T^i
\hat{n}_i$, where $i$ is the calorimeter tower number for $|\eta| <$ 3.6,
 and $\hat{n}_i$ is a unit vector
perpendicular to the beam axis and pointing at the $i^{th}$
tower. We correct $\metvec$ for jets and muons. We define
the magnitude $\met=|\metvec|$.
\end{thebibliography}
\end{document}